\documentstyle[preprint,revtex]{aps}
\begin{document}
\draft
\preprint{UIOWA-94-10}
\begin{title}
Hadron collider limits on anomalous $WW\gamma$
couplings\\
\end{title}
\author{Kevin R. Barger and M. H. Reno}
\begin{instit}
Department of Physics and Astronomy,\\
University of Iowa, Iowa City, IA\ 52242, USA
\end{instit}

\begin{abstract}
A next-to-leading log calculation of the reactions $pp$ and $p\overline{p}$
$\rightarrow W^\pm\gamma X$ is presented
including a tri-boson gauge coupling 
from non-Standard Model contributions.  The additional term 
arises by 
considering the Standard Model as a low energy effective theory.  
Two 
approaches are made for comparison.  The first approach considers the
tri-boson $WW\gamma$ coupling as being uniquely fixed by 
tree level unitarity
at high energies to its Standard Model form and, consequently, 
suppresses
the non-Standard Model contributions with form factors.  The second 
approach
is to ignore such considerations and calculate the contributions to 
non-Standard Model tri-boson gauge couplings without such 
suppressions, 
the idea being that at sufficiently high energies where new physics 
occurs,
one abandons the low energy effective theory.  It is found that at 
Tevatron
energies, the two approaches do not differ much in quantitative
results, while at Large
Hadron Collider (LHC) energies the two approaches give significantly 
different predictions for production rates. At the Tevatron and LHC, 
however,
the sensitivity limits on the anomalous coupling of $WW\gamma$ are 
too weak to
usefully constrain parameters in effective Lagrangian models.
\end{abstract}

\narrowtext
\section{INTRODUCTION}

With the advent of hadron supercolliders, it will be possible to 
directly test the tri-boson couplings of the $W$, $Z$ and photon.
Indeed,
observations of the $WW\gamma$ coupling have 
been reported from measurements of $\sigma(p\overline{p}
\rightarrow W^\pm \gamma)$\cite{alitti,DZERO,CDF}.
Experimental measurements coupled with accurate theoretical predictions
could result in the confirmation of the Standard Model (SM), or
alternatively, could point to 
new physics above the $Z$ mass scale.  The tri-boson couplings are 
uniquely constrained in the Standard Model if the gauge symmetry is
${\rm SU}(2)\otimes {\rm U}(1)$, the symmetry breaking sector is given
by a minimally coupled, scalar Higgs boson, and the theory is
renormalizable.
If the Higgs boson as described by the SM is not the final word,
the possibility exists that the SM is a low energy approximation
to a more fundamental theory.  In this case, 
non-Standard Model (NSM) effects may modify the tri-boson couplings.

The specific process $p\bar{p}$ or $pp$ production of $W^\pm\gamma$
is of particular interest, in part because of the presence of
an amplitude zero.
In the parton  center of mass frame, a zero in the amplitude occurs
at a fixed angle between the quark and photon 
momenta\cite{ZERO,NEWZERO}. 
In moving from the parton center of
mass frame to the hadron collider frame, the zero translates to
a dip in the angular distribution of the photon.
Anomalous $WW\gamma$ couplings result in
contributions that fill in the zero, making the $W^\pm\gamma$
production process especially sensitive 
to nonstandard effects at leading
order. This has been explored in the literature by a variety
of authors\cite{FLS,CORTES,ZEPP,BERGER}. 
Some of these authors treat
the anomalous $WW\gamma$ couplings as constants\cite{FLS}  
while others
include form factors multiplying the NSM 
parameters\cite{CORTES,ZEPP,BERGER}.

It has also been shown that strong interaction 
corrections\cite{THOMAS,JJVDB,ERREDE,OHNEMUS,BAUR,SMITH}
fill in the dip in the photon angular distribution.
Thus, the strong interaction corrections must be well understood 
before one can claim evidence of new physics.
The next-to-leading order calculation of $W\gamma$ 
production, including
non-standard couplings, has been performed by Baur, Han and Ohnemus
in Ref. \cite{BAUR}.
However, their estimation of the sensitivities of the 
Fermilab Tevatron and supercolliders to
nonstandard couplings
was done using form factors.
In the effective Lagrangian approach of Ref. \cite{FLS}, however, 
there are no form factors multiplying the NSM parameters, 
but neither are strong interaction corrections included.
At leading order and next-to-leading order in QCD, the 
two approaches to incorporating
nonstandard couplings (with and without form factor suppression) 
lead to significant differences in $W^\pm\gamma$ production
rates for the Large Hadron Collider (LHC). 
The focus of this paper is the comparison of the
two approaches, including the $O(\alpha_s )$ corrections, 
to $W^\pm \gamma$
production at the Tevatron and LHC.

The rest of this paper is organized as follows:
In Section II, brief descriptions of 
nonstandard couplings and the Monte Carlo method incorporating
the QCD corrections are included.
Experimental cuts and approximations of the method are described 
along with
some of the theoretical uncertainties.  In Section III, the results 
are presented.  Conclusions are presented in Section IV.  

\section{CALCULATION}

\subsection{Non-Standard Couplings}

The calculations of Baur, Han and Ohnemus in Ref. \cite{BAUR} 
are based on a phenomenological Lagrangian,
\begin{equation} 
{\cal L }_{WW\gamma} = 
 -{\rm i} e \biggl( W^{\dagger}_{\mu\nu} W^\mu A^\nu 
-  W^{\dagger}_\mu A_\nu W^{\mu\nu}  + \kappa W^\dagger_\mu
W_\nu F^{\mu\nu} + {\lambda \over M^2_W} W^\dagger_{\lambda\mu}
W^\mu_\nu F^{\nu\lambda} \biggr)\label{eq:tpo}
\end{equation}
where $A^\mu$ and $W^\mu$ are the photon and $W$ fields, 
respectively.
The field strength tensors have the usual definitions for Abelian
and non-Abelian gauge fields.
In the standard model, $\kappa = 1$ and is related
to the $W$ magnetic dipole moment $\mu$ and electric quadrupole moment
$Q$ by  
\begin{equation}
 \mu = (1+\kappa){e\over 2M_W},\>\>\>
Q = - \kappa{2e\over M^2_W}.
\end{equation}
The standard model value for $\lambda$ is $\lambda=0$.
In Ref. \cite{BAUR} , the NSM calculation is performed with 
$\Delta\kappa = 
\kappa-1$ and $\lambda$ of 
Eq. (\ref{eq:tpo}) scaling as
\begin{equation}
 \Delta\kappa(M^2_{W\gamma},p^2_W=M^2_W,p^2_\gamma=0) =
{\Delta\kappa_0 \over ( 1+ M^2_{W\gamma}/\Lambda^2)^n}.
\label{eq:formfactor}
\end{equation}
\begin{equation}
\lambda(M^2_{W\gamma},p^2_W=M^2_W,p^2_\gamma=0) =
{\lambda_0 \over (1+M^2_{W\gamma}/\Lambda^2)^n},
\end{equation}
to preserve unitarity at asymptotically high energies. 
Here $\Lambda$ is
the scale at which new physics becomes important,  $M^2_{W\gamma}$ 
is the invariant mass of the $W\gamma$ system, and $\lambda_0$ and
$\kappa_0$ are the coupling parameters at low energy appearing in 
Eq. (\ref{eq:tpo}). 
In what follows, we drop the subscripts on $\kappa$ and $\lambda$,
and we indicate explicitly when the form factor is included.

Presumably $\Lambda \gg
M_Z$ where $M_Z$ is the $Z$ mass.  In the calculation
presented here, $\Lambda$ is taken as 1 TeV and $n=2$ in
the form factors, as in Ref. \cite{BAUR}.
These form factors correspond to dipole factors similar to the roles
played by the nucleon electric and magnetic dipole
form factors appearing in deep inelastic
scattering experiments.  In those experiments\cite{NUCFORM}, it is 
found empirically that
the nucleon form factors have an approximate dipole form at low energies.
There is, however, no {\it a priori} reason to expect the same here.

Longhitano has demonstrated in Ref. \cite{LONGHITANO} that the most 
general CP conserving, dimension four operators which preserves the
${\rm SU}(2) \otimes {\rm U}(1)$ symmetry in the 
effective Lagrangian approach for the $WW\gamma$ vertex 
leads to 
the standard model terms, plus a term with
$\Delta\kappa\neq 0$. 
In what follows,
the parameter $\lambda$ of Eq. (\ref{eq:tpo}) 
is ignored and the only NSM parameter
of concern is $\Delta\kappa$.  
The notation of Ref. \cite{FLS} has
$\Delta\kappa$ is written in terms of the parameter $\hat{x}$:
\begin{equation} 
\Delta\kappa =  { e^2\hat{x} \over 16\pi^2 s^2_Z}.
\end{equation}
Here,
\begin{equation}
s_Z^2 c_Z^2 = {\pi \alpha_{em} \over \sqrt 2 G_F M^2_Z},
\end{equation}
where $\alpha_{em} = e^2/4\pi$ is the electromagnetic fine 
structure constant. 
We vary $\hat{x}$ between zero and $\hat{x}=400$, with and without
the form factor suppression of Eq. (\ref{eq:formfactor}).
For convenience of comparison with the literature, the values of 
${\hat x}$ used in this paper translate as follows:
\begin{eqnarray}
{\hat x } &= 50 \Rightarrow \Delta\kappa = 0.13\nonumber\\
{\hat x } &= 200 \Rightarrow \Delta\kappa = 0.53\nonumber\\
{\hat x } &= 400 \Rightarrow \Delta\kappa = 1.06\ .
\label{eq:xhatvalues}
\end{eqnarray}

Possible C or P violating terms are not included here
as they are constrained by
experimental measurements of the neutron electric dipole moment to be
negligibly small compared to the non-CP violating anomalous 
couplings\cite{CPVIOL}.  

The Feynman rules for the $WW\gamma$ vertex from the Lagrangian of
Eq. (\ref{eq:tpo}), with the momentum assignments 
$W_\beta(q)\rightarrow W_\mu(p_3)+\gamma_\nu(p_4)$, 
and setting $\lambda = 0$, give
\begin{eqnarray}
\Gamma^{\rm SM}_{\beta\mu\nu} (q=p_3+p_4,p_3,p_4)
=& -{\rm i}e\, Q_W
\biggl(g_{\beta\mu}(q+p_3)_\nu\nonumber\\
&\ -g_{\beta\nu}(q+p_4)_\mu
+g_{\mu\nu}(p_4-p_3)_\beta\biggr)
\end{eqnarray}
\begin{equation}
\Gamma^{\rm NSM}_{\beta\mu\nu}(q=p_3+p_4,p_3,p_4) =  -{\rm i}e\,Q_W
(\Delta\kappa ) \biggl( g_{\beta\nu} p_{4\mu} - 
g_{\mu\nu} p_{4\beta}\biggr)
,
\end{equation} 
where $Q_W=\pm 1$ is the $W$ charge.

\subsection{Methodology}

The calculations described here are the Born level,
bremsstrahlung and $O(\alpha_s )$ corrections
to $p\overline{p}$ and $pp$ production of $W^\pm \gamma X$, 
to yield parton level
results for $q_1\bar{q}_2\rightarrow W^\pm \gamma$, 
$q_1\bar{q}_2\rightarrow W^\pm \gamma g$ and 
$g q_1 (\bar{q}_2)\rightarrow
W^\pm g q_2 (\bar{q}_1)$, in terms of a parton level differential
cross section $d\hat{\sigma}$. 
The leading logarithm (LL) result includes
the Born and bremsstrahlung contributions. The next-to-leading 
logarithmic (NLL) contributions include the interference of
Born and virtual diagrams, the $O(\alpha_s)$
tree level corrections and NLL quark fragmentation into a photon.
The parton differential cross sections
are convoluted with the parton
distribution functions $F^A_i(x_i, Q^2)$ 
for parton $i$ and hadron $A$, 
and summed to yield the differential cross
section 
\begin{equation}
{\rm d}\sigma = \sum_{i,j} \int \int F^{(A)}_i\left(x_1,Q^2\right)
F^{(B)}_j\left(x_2,Q^2\right) 
d\hat\sigma_{ij}\left(\alpha_s\left(\mu^2\right),Q^2\right) 
{\rm d}x_1 
{\rm d}x_2 \ .
\end{equation}
For the
evaluation of the above integral, a combination of analytic
and Monte Carlo techniques is employed.
This method has been used in several processes, including
the $W^\pm\gamma X$ process\cite{OHNEMUS,BAUR,SMITH}.

First, the collinear and soft divergences in the phase
space are isolated by partitioning phase space
with arbitrary (but small) cutoff parameters $\delta_s$ 
and $\delta_c$. Next, the phase space integrals are 
performed analytically in $N=4-2\epsilon$ 
dimensions where the collinear and soft gluon singularities appear as
poles in $\epsilon$.  In the singular regions, 
the three body matrix elements
are approximated using the soft gluon and leading pole 
approximations\cite{BERGMANN}.
With the singularities of the two and three body matrix 
elements isolated, the soft singularities cancel the 
virtual singularities arising from
loop integrals performed in calculating the virtual corrections.
The collinear singularities are then factorized into
the parton distribution functions.
The Born, virtual, soft and collinear contributions  are combined into
what are called the two body
matrix elements, since, at least approximately, they all have the same
$2\rightarrow 2$ kinematics.  
The tree level, non-singular contribution has $2\rightarrow 3$
kinematics and is referred to as the three body contribution.
The two  and three body contributions are singularity free and 
may be integrated via standard Monte
Carlo techniques.  The separate two and three body contributions 
now depend on the theoretical soft and collinear cutoffs 
$\delta_s$ and $\delta_c$, but when the two and three body 
contributions are combined, this dependence vanishes 
for a wide range of cutoff parameters.

The expressions for the two body
matrix elements 
for the SM Lagrangian
have been given in Ref. \cite{OHNEMUS} 
and will not be given here. We do include
in the Appendix the virtual corrections that involve
the non-standard tri-boson
coupling.
The three body matrix elements are calculated 
with the helicity amplitude 
method detailed in Ref. \cite{BHZ} and references contained therein.

Photon bremsstrahlung contributions from final state radiation
from a quark or antiquark are included via the inclusion of the NLL
fragmentation functions for $q\rightarrow \gamma$ and 
$g\rightarrow \gamma$
as in Refs. \cite{OHNEMUS}
and \cite{STIRLING}. A photon isolation cut requires that the sum of
hadronic energy within a cone around the photon momentum be small.
Quantitatively, 
\begin{equation}
\sum_{\Delta R\leq 0.4} E_{had}<0.15 E_\gamma \label{eq:isolation}
\end{equation}
effectively suppresses the bremsstrahlung cross section\cite{STIRLING}.
Here, $\Delta R=[(\Delta\phi )^2+(\Delta \eta )^2]^{1/2}$ 
is the cone size
defined with respect to the photon pseudorapidity
$\eta$ and azimuthal angle $\phi$.

Photons can arise from radiative $W$ decay as well as from the
production process. The signature of $W\rightarrow e\nu_e\gamma$
can, in principle, be separated from the production process 
experimentally by
suitable kinematic cuts.  In particular if a cut, 
$m_{\rm T}(\gamma \ell;{\rm missing}) >  90\ {\rm Gev}$,
is made on the cluster transverse mass variable
\begin{equation}
m_{\rm T}^2(\gamma \ell;{\rm missing}) = \biggl[m^2_{\gamma \ell} + 
\mid {\bf p}_{\gamma {\rm T}}+ {\bf p}_{\ell {\rm T}}\mid^2 +
\mid{\bf p\llap/}_{\rm T}\mid\biggr]^2 
-\mid{\bf p}_{\gamma {\rm T}}+{\bf p}_{\ell{\rm T}}+
{\bf p\llap/}_{\rm T}\mid^2 
\end{equation}
most of the radiative decay signal will be eliminated \cite{CORTES}.
Here ${\bf p\llap/}_{\rm T}$ 
is the neutrino transverse momentum, which is
missing in an experiment, and ${\bf p}_{\ell {\rm T}}$
is the lepton transverse momentum.
In this calculation only the production process $p\overline{p}$ or $pp$ 
$\rightarrow W^\pm\gamma X$ is considered for simplicity.  
The decay processes into leptons $\ell =e$ or $\ell = \mu$:
$W \rightarrow {\ell}\nu_{ \ell}$ and $W \rightarrow {\ell}\nu_{ \ell}
\gamma $ are
ignored throughout, except as the branching fraction $BR(W\rightarrow
\ell \nu_\ell )$ appears in event rates.

The strong coupling constant $\alpha_s$ is calculated at two loops
for the NLL results and at one loop for the LL results,
with the five light quark flavors contributing at their respective mass
thresholds.  
Also, the HMRS set B  structure functions\cite{HMRS} consistent
with a NLL, $\rm\overline{MS}$ (Modified Minimal Subtraction scheme) 
calculation are used.  For these structure functions, the four flavor
value of $\Lambda_{{\rm \overline{MS}}}$ of 0.19 GeV is used.
For convenience, we set the factorization scale equal to the
renormalization scale. Unless otherwise specified, these scales are
set to the $W\gamma$ invariant mass $M_{W\gamma}$.
The electromagnetic fine structure constant at the $Z$ mass scale 
$\alpha_{em} = 1/128.8$ is used.
The narrow width approximation for the $W$ propagator
\begin{equation}
{1\over (s-M^2_W)^2 + M^2_W\Gamma^2_W} \approx {\pi\over M_W \Gamma_W}
\delta(s-M^2_W)\>\>\>
\end{equation}
is also used.

In addition to the photon isolation cut described above,
various kinematic cuts are imposed on the variables to simulate detector
responses.  The $W$ and photon are both required 
to lie in a rapidity range of $| y | \le 2.5$.
A cut is made on the photon transverse momentum to avoid the collinear and
soft singularities arising from the parton level cross section.  For the
Tevatron,  we select photons with $p_{\gamma\perp }
> 10$ GeV, while at the LHC, we 
require
$p_{\gamma\perp } > 50$ GeV.  

\vfil\eject
\section{RESULTS}

The calculation described in Sec. II 
has been performed for $\sqrt S = 1.8$
TeV (Tevatron) and $\sqrt S = 14$ TeV (LHC) 
for $p\overline{p}$ and $pp$, respectively.  
In the
tables and figures presented below, the number of events, $N$, 
is obtained from 
\begin{equation} 
N = {\cal L} \times BR \times 
\int_{p_{1\gamma{\perp}}}^{p_{2\gamma{\perp}}}
{{\rm d}\sigma \over {\rm d} p_{\gamma\perp }}{\rm d}p_{\gamma\perp }
\end{equation}
where $BR\approx 0.2$ is the sum of the
electronic and muonic branching fractions for the $W$, 
${\cal L}$ is the integrated luminosity ${\cal L} = \int L dt$
over a collider year,  
and ${{\rm d}\sigma / {\rm d} p_{\gamma\perp}}$
is the $| {\bf p}_{\gamma\perp}|$ spectrum of the photon.  
The calculations are performed
with various values of $\hat x = 0,\>50,\>200,\>400$, and with
different regions of $p_{\gamma\perp }$.  The value of
$\hat x = 0$ corresponds to the SM.
The integrated luminosity
for the Tevatron is taken as ${\cal L} = 100\>\>{\rm pb}^{-1}$,  
while the
LHC luminosity used is ${\cal L} = 3\times10^4\>{\rm pb}^{-1}$. 
For the $pp$ collider the event
rates for $W^+$ are slightly higher than for $W^-$, while at the
$p \bar p$ collider the event rates for $W^\pm$ are the same.  

\subsection{Tevatron Collider}

We begin with a comparison of 
the LL event rates along with the NLL values.  These calculations
for the Tevatron Collider are presented in Tables 1 and 2.
The trends
are evident in the tables.  
At the Tevatron, with no form factors, 
the LL and NLL event rates are enhanced
with increasing $\hat x$.  The amount of increase is small, however,
for low $p_{\gamma\perp }$. 
At the leading log level, the difference from ${\hat x} = 0$ to 
${\hat x} = 400$ is only 14 events at $10\>\>{\rm GeV} 
\le\> p_{\gamma\perp }  \>\le 50\>\>{\rm GeV}$.  This 
corresponds to a 10\% increase.  At $50\>\>{\rm GeV} 
\le\>p_{\gamma\perp } \>\le 150\>\>{\rm GeV}$, the increase is from
6 to 14 events.  
The SM signal is enhanced by about 25\% when
the NLL corrections are included at the low $p_{\gamma\perp }$ end.  
At the
high $p_{\gamma\perp }$ end, the increase is 50\%, but this only 
means an
increase of 3 events, not statistically significant given the 
associated statistical uncertainty. 
All of this can be seen qualitatively by examining the photon
transverse momentum spectra shown in Figs. \ref{fthree} (LL)  and 
\ref{ffour} (NLL), where we
show ${\rm d}\sigma(p\bar{p}\rightarrow W^+\gamma X)/
{\rm d}p_{\gamma\perp }$ 
at
the Tevatron. The anomalous $WW\gamma$ coupling enhances the high
$p_{\gamma\perp }$ tail of the distributions.
No form factors are used for the distributions. 

Tables
1 and 2 indicate that   
the event rates at the Tevatron are nearly equal with and without form
factors, for both LL and NLL results.  
This is due to the fact that
$M_{W\gamma}^2/\Lambda^2$ in the form factor is small.
The differences in number of events 
are not statistically significant, although
the form factor approach does decrease the number of events. Thus,
limits obtained including the 
form factor suppression could be considered
to be conservative.

The D0 Collaboration reports 
a value of $-2.5\leq \Delta\kappa \leq 2.7$\cite{DZERO} 
from a data sample of
approximately 15 pb$^{-1}$ from the 1992/93 run.
The limits are set using the NSM enhancement in the total cross
section. To estimate the sensitivity at an integrated luminosity of
100 pb$^{-1}$, 
we consider the high $p_{\gamma\perp }$ region in Table 2.
We see from
Table 2 that a doubling of the SM event rate from 9 to 18 events for 
50 GeV$\leq p_{\gamma\perp } \leq$150 GeV
occurs at $\hat x \simeq 400$.  For comparable values of $\hat{x}$, 
it would be
difficult to set a limit from the cross section alone because
the NSM enhancement of the cross section is not very large, less than
10\%\ for $\hat{x}=400$.
From Eq. (\ref{eq:xhatvalues}),  this corresponds to 
$\Delta\kappa \simeq 1$. 

Baur, Han and Ohnemus in Ref. \cite{BAUR} have estimated a sensitivity
to $|\Delta\kappa|\simeq 1.6$ at the 2$\sigma$ level and
$|\Delta\kappa|\simeq 0.9$ at the 1$\sigma$ level for
an integrated luminosity of
100 pb$^{-1}$ at the Tevatron. Their results for positive and negative
$\Delta\kappa$ are approximately equal.
They take
$p_{\gamma\perp } >10$ GeV, 
and impose a variety of other cuts including those on the
leptons from the $W$ decay.
From Tables 1 and 2 combined, we see that the enhancement of the
NLL over the LL results is approximately constant as a function of
$\hat{x}$ when one includes $p_{\gamma\perp }
>10$ GeV. Since the form factor has
at most a few percent effect in the direction of suppressing
the nonstandard contributions, the Tevatron
results of Ref. \cite{BAUR} can be
carried over to the case where no form factors are used.
A value of $\Delta\kappa=1.6$ corresponds to $\hat{x}\simeq 600$.
The limits of Ref. \cite{BAUR} apply to $\Delta\kappa$ for any value of
$\lambda$ in Eq. (\ref{eq:tpo}), 
so with $\lambda=0$, the limits on $\Delta\kappa$ 
improve somewhat, an estimated 10--30\% \cite{ERREDE}.

\subsection{Large Hadron Collider}

At the LHC, the results are somewhat different than
for the Tevatron.  The discussion will
cover only the case of $W^+\gamma$ production, with similar conclusions
for $W^-\gamma$ production. As with the Tevatron Collider, as $\hat{x}$
increases, so does the cross section 
because of a flatter $p_{\gamma\perp }$ 
distribution.
Event rates for $\sqrt{S}=14$ TeV for two ranges of $p_{\gamma\perp }$ 
are shown
in Table 3 (200 GeV$\leq p_{\gamma\perp }
\leq 400$ GeV) and Table 4 (400 GeV$\leq
p_{\gamma\perp }\leq 750$ GeV), using the cuts described in Sec. II.
Our results for large $\hat{x}$ at the LHC 
indicate that the cross section
scales approximately as $\hat{x}^2$. We use this approximate $\hat{x}^2$
dependence to extrapolate between values of $\hat{x}$ given in
Tables 3 and 4. In addition, our conclusions should be valid for
positive and negative $\hat{x}$.
The LHC results, because of the significantly
higher energy at the LHC, exhibit two striking features. First,
as has been pointed out in the literature\cite{OHNEMUS,BAUR,SMITH},
the QCD corrections are enormous. Second, the form factor results are
measurably lower than the results with no form factor.

Tables 3 and 4 demonstrate that the QCD corrections overwhelm 
the SM LL signal at the
LHC, being as much as five times the LL signal at low $p_{\gamma\perp }$,
and increasing to seven times the LL signal at 
the high $p_{\gamma\perp }$ 
end of
the spectrum in the SM. 
These large contributions at ${\cal O}(\alpha_s)$
would seem to cast doubt on the validity of the perturbative
expansion, but are generally caused by the presence of the radiation
zero mentioned above, thereby suppressing the photon transverse 
momentum
distributions at LL order\cite{OHNEMUS}.
The transverse momentum distributions at
LL and NLL for $pp\rightarrow W^+\gamma X$ at the LHC, without
form factor suppression, are shown
in Figs. \ref{fnine} and \ref{ften}.  
Ref. \cite{BAUR} describes a method to reduce 
QCD effects in $W\gamma$ production 
by putting appropriate cuts on the final
state parton that appears in the $O(\alpha_s )$ tree level
matrix element.  This effectively
reduces the large contributions from the ${\cal O}(\alpha_s)$ 
corrections, but also reduces the event rate.
Such cuts will not be considered here.
Instead, we consider the quantity 
\begin{equation}
{\sigma(p_{\gamma\perp } > p_{\gamma\perp {\rm min}}) \over
\sigma(p_{\gamma\perp } > 50\ {\rm GeV}) }\ ,
\end{equation}
where some of the QCD uncertainties cancel in the ratio.

In Fig. \ref{fthirteen}, 
we show the cross section ratios for the standard model 
at LL and NLL and the NSM results at NLL for 
$\hat{x}=50$, 200 and 400. The
ratio reflects the significant increase 
in events at high $p_{\gamma\perp }$ due to
the QCD and NSM effects seen in Figs. \ref{fnine} and \ref{ften}.
We have put in error bars to indicate 
the statistical errors given an 
integrated luminosity of $3\times 10^4$ pb$^{-1}$, 
including the branching
fractions for $W\rightarrow e$ and $W\rightarrow \mu$. 
The ${\hat x} = 50$ signal is
a 1$\sigma$ distance from the SM for the $p_{\gamma\perp }
\buildrel >\over\sim 400$ GeV.
For Fig. \ref{fthirteen},
the factorization and renormalization scales $Q$ are kept at a value
of $Q=M_{W\gamma}$. 
Fig. \ref{ffifteen} shows the cross section ratios as a
function of $p_{\gamma\perp {\rm min}}$, at NLL, 
for three values of $Q$.
The spread in the ratio stays within the statistical error bars, 
but the QCD uncertainty is
large enough to make a limit of $\hat{x}\sim 50$ difficult without
additional cuts invoked\cite{NOTE}. 

The analysis of Falk, Luke and Simmons\cite{FLS} of the sensitivity of
the LHC to $\hat{x}\neq 0$ used the number of excess
events in the high $p_{\gamma\perp }$ range, at leading order,
as a guide to LHC sensitivity. 
They required a doubling of the standard model event rate.
Table 4 indicates that the 
LHC would
be sensitive to $\hat{x}\simeq 50$ 
at leading order with this criterion.
This is roughly the conclusion of Ref. \cite{FLS} in their
$W\gamma$ discussion for a higher LHC energy 
but lower integrated luminosity.
At NLL, however, the same criterion 
means a sensitivity to $\hat{x}\sim 140$. 
An alternative way to assess the importance
of the QCD corrections is to compare the standard model NLL rate 
with the nonstandard model LL rate. The number of predicted events
for $400\ {\rm GeV}\leq p_{\gamma\perp }
\leq 700$ GeV at the LHC is equivalent, 
at LL, to a value of $\hat{x}\sim 140$. 

We now turn to the issue of form factor suppression of NSM effects.
Tables 3 and 4, and
Figs. \ref{feleven} and \ref{ftwelve} 
indicate the degree to which the form factor suppresses
the cross section.
Figs. \ref{feleven} 
and \ref{ftwelve} show the LL and NLL 
$p_{\gamma\perp }$ distributions. In
these figures as well as the form factor results in the tables,
the form factor of Eq. (\ref{eq:formfactor}) 
is used with $\Lambda=1$ TeV and $n=2$.
With form factors applied, the increases in Table 3 from
the SM are a bit more modest than without the form factor:
the SM LL event rate increases roughly by a factor of five 
for $\hat{x}=400$, and the NLL increase is less than twice the SM
at the same value of $\hat{x}$.
At high $p_{\gamma\perp }$,
$400\>\>{\rm GeV} \le\> p_{\gamma\perp } 
\>\le 700\>\>{\rm GeV}$ with no form factors
applied the increases from the SM values for LL and NLL are
sixty and ten respectively for $\hat{x}=400$.  
With form factors applied,  the increase from $\hat{x}=0$ to
$\hat{x}=400$ for LL is
about five times the SM value.
At NLL the increase from SM and  
$\hat x = 400$ is a relatively modest 17\%.

Because the form factor depends on $M_{W\gamma}$, 
the translation between
results with and without form factors is not completely 
straightforward.
From Table 3, where 200 GeV$\leq p_{\gamma\perp }
\leq 400$ GeV, the number of events
with the form factor applied at $\hat{x}=400$ is equivalent 
to the number
of events predicted without form factors at $\hat{x}\simeq 200$. 
In this range of $p_{\gamma\perp }$, the average value of 
$M_{W\gamma}$ is 
$<M_{W\gamma}>\sim$600 GeV, which accounts for the suppression.
At larger values of $p_{\gamma\perp }$, 
the suppression is stronger. In
Table 4, the event rate for $\hat{x}=400$ with the
form factor is equivalent to $\hat{x}\simeq 150$ without
form factor suppression.
At lower values of $p_{\gamma\perp }$, 
the form factor is less important. For
100 GeV$\leq p_{\gamma\perp }\leq 200$ GeV, 
$<M_{W\gamma}>\simeq 330$ GeV.
From this, we estimate that $\hat{x}=400$ with 
form factor multiplication
is equivalent to $\hat{x}\simeq 325$ in this $p_{\gamma\perp }$ 
bin, effectively a
factor of 0.8 lower in the $\hat{x}$ value.

We use these comparisons to translate the sensitivity limits of 
Baur, Han
and Ohnemus in Ref. \cite{BAUR}  to limits without form factors.
In Ref. \cite{BAUR}, 
the inclusive NLL distributions are used to set sensitivity
limits of $\Delta\kappa\sim 0.3$ at 
1$\sigma$ and $\Delta\kappa\sim 0.5$
at 2$\sigma$ at $\sqrt{S}=40$ TeV for the 
Superconducting Super Collider,
for any value of $\lambda$ in Eq. (\ref{eq:tpo}). 
They comment that the LHC limits are larger by a 
factor of $\sim 1.5$.
In terms of $\hat{x}$, this means a sensitivity 
(with form factors) of
$\hat{x}\sim 180 - 300$ for the 1--2$\sigma$ range. 
Their analysis involves
a variety of cuts, including cuts on the leptons as well as
$p_{\gamma\perp }\geq 100$ GeV. Since low $p_{\gamma\perp }$ 
dominates
the cross section, a conservative estimate, 
using the multiplicative factor
of 0.8, would be that the LHC is sensitive
to $\hat{x}\sim 150-250$ in the absence of 
form factors in the calculation.
A less conservative estimate would be to take the factor of 
0.5 found
from the range of 200--400 GeV for $p_{\gamma\perp }$ 
to yield 90--150 for $\hat{x}$ when
form factors are taken out of the analysis.

We comment that Baur {\it et al.} have also considered the 
next-to-leading
order $W\gamma+0$-jet rate, as a way to eliminate some of the QCD 
uncertainties. The inclusive NLL limits are a factor of 
1.2-1.5 higher
than those obtained with the no-jet rate\cite{BAUR}.
In addition, they use an integrated luminosity of 
$10^4$ pb$^{-1}$ and only include
the $W^+$ production with $W\rightarrow e\nu_e$ decay channel, and
comment that the limits can be improved by 20--40\%\ with the 
inclusion of $W^-$ and $W\rightarrow \mu\nu_\mu$. 
By accounting for these improvements
and an integrated luminosity of $3\times 10^4$ pb$^{-1}$, 
one is led to a lower limit on $\hat{x}$:
at best $\hat{x}\sim 150-250$ is reduced to $\hat{x}\sim 80-140$.

\section{CONCLUSIONS}

A comparison 
sensitivity limits with and without form factors,
including QCD corrections, has been done for
the NSM parameter $\hat{x}>0$,  which also 
applies to negative values of $\hat{x}$.
We have demonstrated that QCD corrections and form factor suppression
have significantly different effects at Tevatron and LHC energies.
At the Tevatron, QCD corrections are at the level of $\sim 35\%$ 
of the Born cross section. The
form factor suppression 
at the Tevatron is essentially negligible, so
the results of Ref. \cite{BAUR}, 
which include QCD corrections, can be
carried over to the effective Lagrangian approach. Their estimated
sensitivity to the nonstandard coupling is at $\hat{x}\sim 360-600$,
the range of $\Delta\kappa \simeq 0.9$ at the 1$\sigma$ level and
$\Delta\kappa\simeq 1.6$ at the 2$\sigma$ level. 
Our cruder estimate,
requiring a doubling of the SM events for 50 GeV 
$\leq p_{\gamma\perp }\leq$ 150
GeV, yields a comparable value of $\hat{x}=400$. 

At the LHC, however, the QCD corrections play 
a much more important role
in the sensitivity limits. Doubling the number of events at LL,
for 400 GeV$\leq 
p_{\gamma\perp }\leq$ 700 GeV occurs for $\hat{x}\simeq 50$,
while at NLL, for the same range of $p_{\gamma\perp }$, 
a value of $\hat{x}\simeq 140$
is required. The more complete analysis of Ref. \cite{BAUR}, 
done with form factors and a variety of theoretical cuts, 
can be translated to effective Lagrangian results by
evaluating $<M_{W\gamma}>$. We estimate that
the limits of Ref. \cite{BAUR} translate to a 
sensitivity to $\hat{x}\simeq
150-250$ without form factors, although it is possible that 
these limits could be reduced somewhat. 

Each collider energy has its advantages, 
however, the conclusion that we
must draw is that neither the Tevatron with 
${\cal L}=100$ pb$^{-1}$ 
nor the LHC with ${\cal L}=3\times 10^4$ pb$^{-1}$ will be 
sensitive to values of the anomalous $WW\gamma$ coupling 
$\hat{x}$ that are relevant
to the effective Lagrangian approach. 
With the definition of $\hat{x}$ in
Eq. (2.5), the value of $\hat{x}$ in the effective
Lagrangian is naively expected to be of order unity\cite{FLS}. 
If values of $\hat{x}\sim 100$ were measured in experiments, 
one would bring into question the
validity of the effective Lagrangian approach where only 
one nonstandard
coupling is kept in the $WW\gamma$ effective Lagrangian.
This is the same conclusion for the $WW\gamma$ anomalous coupling
reached in Ref. \cite{FLS} with a leading order analysis. 
Our results are more pessimistic, 
because using the same criterion, a doubling of events at 
high $p_{\gamma\perp }$,
the inclusion of QCD effects weakens the estimate in 
Ref. \cite{FLS} of
the LHC sensitivity to $\hat{x}$ by
a factor of $2-3$.

\acknowledgements

We thank U. Baur for discussions.
This work was supported in part by NSF Grants No. PHY-9104773 and No.
PHY-9307213.  

\unletteredappendix{VIRTUAL APPENDIX}

The matrix element squared for the one loop virtual correction
to $q_1(p_1)+\bar{q}_2(p_2)\rightarrow W(p_3)+\gamma(p_4)$ is
written in terms of Mandelstam invariants
\begin{equation}
\matrix{s=(p_1+p_2)^2 & t=(p_1-p_4)^2 & u=(p_2-p_4)^2 }\ . 
\end{equation}
It has the form
\begin{eqnarray}
\sum_{avg} |M_V|^2 &=&
{1\over 9}\, {1\over 4} {e^4\over 2\sin^2\theta_W}
N_C  {\alpha_s\over 2\pi}\Biggl({4\pi\mu^2\over M_W^2}\Biggr)
^\epsilon{\Gamma (1-\epsilon )\over \Gamma (1-2\epsilon )} C_F 
\nonumber\\
& & \nonumber\\
&\cdot& \Biggl[ F_V(s,t,u,\epsilon )+G_V(s,t,u,\epsilon )\Biggr]
\end{eqnarray}
$F_V$ is the standard model contribution and it is explicitly written
out in Ref. \cite{OHNEMUS}. We have check that Ohnemus' 
expression is correct.
The NSM contribution to the virtual correction appears in $G_V$, 
which is written in terms of the NSM part of
the Born matrix element squared
\begin{eqnarray}
T_1 &= & {(Q_1-Q_2)^2\over 2M_W^2(t+u)^2}(\Delta\kappa)^2
\Biggl[ s(t+u)^2(1-\epsilon )+4tu(t+u)(1-\epsilon )\nonumber\\
&+& 2stu(1-2\epsilon )\Biggr] \\
T_2 & = & {(Q_1-Q_2)(Q_1u+Q_2t)\over (t+u)^2}4(t-u)(\Delta\kappa ) 
(1-\epsilon)
\end{eqnarray}
We calculate the NSM virtual correction to be
\begin{eqnarray}
G_V(s,t,u,\epsilon )&=&[T_1(s,t,u,\epsilon )+T_2(s,t,u,\epsilon )]
\Bigl( {s\over M_W^2}\Bigr)^{-\epsilon}
\Bigl( -{2\over \epsilon^2}-{3\over \epsilon}+{2\over 3}\pi^2
\Bigr)\nonumber\\
& + &{\cal F}(Q_1,Q_2,s,t,u)+{\cal F}(Q_2,Q_1,s,u,t)
\end{eqnarray}
where
\begin{eqnarray}
{\cal F}(Q_1,Q_2,s,t,u)&=& {Q_1 (Q_1-Q_2)\over t+u}(\Delta\kappa )
\Biggl[4 F_1(s,t)
(M_W^2-t)\nonumber\\   
&+&{\ln^2\Bigl( {s\over M_W^2}\Bigr)} \Bigl(3u+t-{1\over M_W^2}
(t^2+u^2+ut) \Bigr)\nonumber\\
&+&4 {\ln\Bigl( {|t|\over M_W^2}\Bigr)} 
 {u\over 2}\Bigl( 2-{t+u\over s+u}
-{M_W^2 u\over (s+u)^2}+{3t\over s+u} \Bigr)\nonumber\\
&+&2 u\Bigl(-{u\over s+u}+7-10 {t\over t+u}\Bigr)\nonumber\\
&+&4(\Delta\kappa)
\Bigl(-{s(t+u)\over M_W^2}-2tu({1\over t+u}+{1\over M_W^2})\Bigr)
\nonumber\\
&+&2 {\ln\Bigl( {s\over M_W^2}\Bigr)}\Bigl( M_W^2-t+{2M_W^2 u\over t+u} 
\Bigr) \Biggr]
\nonumber\\
&+&{(Q_1u+Q_2t)(Q_1-Q_2)\over (t+u)^2}(\Delta\kappa )
\Biggl[  11-\ln^2\Bigl( {s\over M_W^2}\Bigr)\Biggr]
 (u-t)\nonumber\\
&-&{(Q_1-Q_2)^2}{1\over 2}{\ln^2\Bigl( {s\over M_W^2}\Bigr)}
(\Delta\kappa )\Biggl[ 1+{4tu\over
(t+u)^2}-{u^2+t^2+ut\over M_W^2(t+u)}\Biggr]
\end{eqnarray}
with the function $F_1$ defined as
\begin{eqnarray}
F_1(s,t)&=&\ln\Bigl( {s\over M_W^2}\Bigr)\ln\Bigl( {t\over M_W^2-s}
\Bigr)
+{1\over 2}\ln^2\Bigl({M_W^2\over s}\Bigr)
-{1\over 2}\ln^2\Bigl({M_W^2-t\over M_W^2}\Bigr)\nonumber\\
&+&{\rm Li}_2\Bigl({M_W^2
\over s}\Bigr) -{\rm Li}_2\Bigl({M_W^2\over M_W^2-t}\Bigr)\ .
\end{eqnarray}

\vfil\eject
\baselineskip=11pt
\newcount\tn
\tn=1
\hphantom{Table \number\tn.}
\vskip 0.3truein
\halign to \hsize{#\hfil\cr
~~~~~~~~~~~~Table \number\tn.~~Events for ${\sqrt S} = 1.8$ TeV ; $10$ 
GeV  $\le\> \mid {\bf p}_{\gamma\perp}\mid\> \le 50$ GeV \hfil\cr
~~~~~~~~~~~~LL = Leading Log ; NFF = No Form Factors ; FF = Form Factors 
\hfil\cr}
\tabskip=1em plus2em minus.5em
\halign to \hsize{#\hfil&#\hfil&#\hfil&#\hfil&#\cr
\multispan5\hrulefill\cr
\noalign{\bigskip}
& SM & $\hat x = 50$ & $\hat x = 200$ & $\hat x = 400$ \cr
\multispan5\hrulefill\cr
\noalign{\bigskip}
$W^+$(LL,NFF) & 142 & 142 & 147 & 156 \cr
$W^+$(LL,FF)  & 142 & 142 & 146 & 156 \cr
\noalign{\bigskip}
\multispan5\hrulefill\cr
\noalign{\bigskip}
$W^+$(NLL,NFF) & 193 & 194 & 200 & 206 \cr
$W^+$(NLL,FF)  & 193 & 190 & 195 & 210 \cr
\noalign{\bigskip}
\multispan5\hrulefill\cr }
\vskip 0.5truein
\advance\tn by 1
\hphantom{Table \number\tn.}
\vskip 0.3truein
\halign to \hsize{#\hfil\cr
~~~~~Table \number\tn.~~Events for ${\sqrt S} = 1.8$ TeV ; $50$ 
GeV  $\le\> \mid {\bf p}_{\gamma\perp}\mid\> \le 150$ GeV \cr
~~~~~LL = Leading Log ; 
NFF = No Form Factors ; FF = Form Factors \hfil\cr}
\tabskip=1em plus2em minus.5em
\halign to \hsize{#\hfil&#\hfil&#\hfil&#\hfil&#\cr
\multispan5\hrulefill\cr
\noalign{\bigskip}
& SM & $\hat x = 50$ & $\hat x = 200$ & $\hat x = 400$ \cr
\multispan5\hrulefill\cr
\noalign{\bigskip}
$W^+$(LL,NFF) & 6 & 6 & 8 & 14\cr
$W^+$(LL,FF)  & 6 & 6 & 8 & 13\cr
\noalign{\bigskip}
\multispan5\hrulefill\cr
\noalign{\bigskip}
$W^+$(NLL,NFF) & 9 & 9 & 11 & 18 \cr
$W^+$(NLL,FF)  & 9 & 9 & 11 & 16 \cr
\noalign{\bigskip}
\multispan5\hrulefill\cr }
\vfill
\eject

\baselineskip=11pt
\newcount\tn
\tn=3
\hphantom{Table \number\tn.}
\vskip 0.3truein
\halign to \hsize{#\hfil\cr
~~~~~~~~~~~~Table \number\tn.~~Events for ${\sqrt S} = 14$ TeV ; $200$ 
GeV  $\le\> \mid {\bf p}_{\gamma\perp}\mid\> \le 400$ GeV \cr
~~~~~~~~~~~~LL = Leading Log ; 
NFF = No Form Factors ; FF = Form Factors \hfil\cr}
\tabskip=1em plus2em minus.5em
\halign to \hsize{#\hfil&#\hfil&#\hfil&#\hfil&#\cr
\multispan5\hrulefill\cr
\noalign{\bigskip}
& SM & $\hat x = 50$ & $\hat x = 200$ & $\hat x = 400$ \cr
\multispan5\hrulefill\cr
\noalign{\bigskip}
$W^+$(LL,NFF) & 398 & 495 & 1788 & 5875\cr
$W^+$(LL,FF)  & 398 & 425 &  773 & 1837\cr
$W^-$(LL,NFF) & 306 & 367 & 1191 & 3786\cr
$W^-$(LL,FF)  & 306 & 325 &  549 & 1250\cr
\noalign{\bigskip}
\multispan5\hrulefill\cr
\noalign{\bigskip}
$W^+$(NLL,NFF) & 1998 & 2007 & 3147 & 7314\cr
$W^+$(NLL,FF)  & 1998 & 1949 & 2117 & 3138\cr
$W^-$(NLL,NFF) & 1364 & 1363 & 2112 & 4786\cr
$W^-$(NLL,FF)  & 1364 & 1338 & 1463 & 2103\cr
\noalign{\bigskip}
\multispan5\hrulefill\cr }
\vskip 0.5truein
\advance\tn by 1
\hphantom{Table \number\tn.}
\vskip 0.3truein
\halign to \hsize{#\hfil\cr
~~~~~Table \number\tn.~~Events for ${\sqrt S} = 14$ TeV ; $400$ 
GeV  $\le\> \mid {\bf p}_{\gamma\perp}\mid\> \le 700$ GeV \cr
~~~~~LL = Leading Log ; NFF = No Form Factors ; 
FF = Form Factors \hfil\cr}
\tabskip=1em plus2em minus.5em
\halign to \hsize{#\hfil&#\hfil&#\hfil&#\hfil&#\cr
\multispan5\hrulefill\cr
\noalign{\bigskip}
& SM & $\hat x = 50$ & $\hat x = 200$ & $\hat x = 400$ \cr
\multispan5\hrulefill\cr
\noalign{\bigskip}
$W^+$(LL,NFF) & 34 & 64 & 498 & 1890\cr
$W^+$(LL,FF)  & 34 & 35 &  59 &  132\cr
$W^-$(LL,NFF) & 21 & 36 & 258 &  969\cr
$W^-$(LL,FF)  & 21 & 22 &  34 &   72\cr
\noalign{\bigskip}
\multispan5\hrulefill\cr
\noalign{\bigskip}
$W^+$(NLL,NFF) & 236 & 258 & 724 & 2258\cr
$W^+$(NLL,FF)  & 236 & 228 & 228 &  283\cr
$W^-$(NLL,NFF) & 130 & 140 & 370 & 1178\cr
$W^-$(NLL,FF)  & 130 & 127 & 122 &  152\cr
\noalign{\bigskip}
\multispan5\hrulefill\cr }
\vfill
\eject
\figure{The LL photon transverse momentum distribution
${\rm d}\sigma(p\bar{p}\rightarrow W^+\gamma X)/{\rm d}
p_{\gamma\perp }$ at $\sqrt{S}=1.8$ TeV
for $\hat{x}=0,\ 50,\ 200$ and 400.  
The figure includes Born term
and bremsstrahlung contributions with the cuts described in Sec. II
of the text. 
No form
factors are used to reduce the cross section.\label{fthree}}
\figure{The NLL photon transverse 
momentum distribution at $\sqrt{S}=1.8$,
as in Fig. \ref{fthree}.\label{ffour}}
\figure{The LL $p_{\gamma\perp }$ 
distributions for $pp\rightarrow W^+\gamma X$ 
at $\sqrt{S}=14$ TeV for $\hat{x}=0,\ 50,\ 200$ and 400.  
The figure includes Born term
and bremsstrahlung contributions with the cuts described in Sec. II
of the text. 
No form
factors are used to reduce the cross section.\label{fnine}}
\figure{Same as in Fig. \ref{fnine}, 
now including NLL contributions.\label{ften}}
\figure{Cross section ratios $\sigma (p_{\gamma\perp } > 
p_{\gamma\perp {\rm min}})/
\sigma(p_{\gamma\perp } > 50\ {\rm GeV})$ 
for the LHC with the cuts described
in Sec. II. Shown are curves for the SM at LL and
NLL results for $\hat{x}=0,\ 50,\ 200,\ 400$. 
The error bars are
an estimate of the statistical errors using a leptonic branching 
fraction $BR=0.20$ and integrated
luminosity ${\cal L}=3\times 10^4$ pb$^{-1}$. 
No form factors are applied.\label{fthirteen}}
\figure{The NLL standard model values for
$\sigma(p_{\gamma\perp } > p_{\gamma\perp {\rm min}}) /
\sigma(p_{\gamma\perp } 
> 50\ {\rm GeV})$ as a function of $p_{\gamma\perp {\rm min}}$, 
for $Q=M_{W\gamma}$,
$Q = {1 \over 2}\sqrt{M_{W\gamma}^2 + p_{W\gamma}^2}$ 
and $Q = {\sqrt{\hat s}}$.\label{ffifteen} }  
\figure{The same LL plots at the
LHC as in Fig. \ref{fnine} but with the form factors
of Eq. (\ref{eq:formfactor})\ 
with $\Lambda = 1$ TeV and $n=2$ applied.\label{feleven}} 
\figure{The same plots as Fig. \ref{feleven}, including
the form factor, now at NLL.\label{ftwelve}}


\begin{references}
\bibitem{alitti}J. Alitti {\it et al.} (UA2 Collaboration), 
Phys. Lett. {\bf B277}, 194 (1992).
\bibitem{DZERO}A. L. Spadafora {\it et al.} (D0 Collaboration), Fermilab
preprint FERMILAB-CONF-94-016-E, to appear in the Proceedings
of the 9th Topical Workshop on $\bar{p} p$ Collider Physics, Tsukuba,
Japan, October 1993.
\bibitem{CDF}Th. M\"uller (CDF Collaboration),
UCLA preprint UCLA-PPH0058-11/93, to appear in the Proceedings of
the Europhysics Conference on High Energy Physics, Marseille, France,
July 1993.
\bibitem{ZERO}K. O. Mikaelian, M. A. Samuel and D. Sahdev, Phys. Rev.
Lett. {\bf 43}, 746, 1979.
\bibitem{NEWZERO}U. Baur, T. Han and J. Ohnemus, Florida State University
preprint FSU-HEP-940307 (March 1994).
\bibitem{FLS} A. Falk, M. Luke and E. Simmons, Nucl. Phys. 
{\bf B365}, 523 (1991).
\bibitem{CORTES}J. Cortes, K. Hagiwara and F. Herzog, Nuc. Phys. 
{\bf B278}, 26 (1986).
\bibitem{ZEPP}U. Baur and D. Zeppenfeld, 
Nucl. Phys. {\bf B308}, 127 (1988).
\bibitem{BERGER}U. Baur and E. Berger, 
Phys. Rev. {\bf D41}, 1476 (1990);
Phys. Rev. {\bf D47}, 4889 (1993).
\bibitem{THOMAS}J. Smith, D. Thomas and W. L. van Neerven, 
Zeit. Phys. {\bf C44}, 267 (1989).
\bibitem{JJVDB}U. Baur, E. W. N. Glover and J. J. van der Bij, 
Nucl. Phys. {\bf B318}, 106 (1989).
\bibitem{ERREDE}U. Baur, S. Errede and J. Ohnemus, 
Phys. Rev. {\bf D48}, 4103 (1993).
\bibitem{OHNEMUS}J. Ohnemus, Phys. Rev. {\bf D47}, 940 (1993).
\bibitem{BAUR}U. Baur, T. Han and J. Ohnemus, 
Phys. Rev. {\bf D48}, 5140 (1993).
\bibitem{SMITH}S. Mendoza and J. Smith, SUNY, Stony Brook preprints
ITP-SB-93-72 and ITP-SB-93-80; 
S. Mendoza, J. Smith and W. L. van Neerven,
Phys. Rev. {\bf D47}, 3913 (1993).
\bibitem{NUCFORM}R. E. Taylor 
{\it Proc. 1975 Int. Symp. on Lepton and Photon 
Interactions at High Energy} (ed. W. T. Kirk).
\bibitem{LONGHITANO}A. Longhitano, 
Nucl. Phys. {\bf B188}, 118 (1981).
\bibitem{CPVIOL}W. J. Marciano and 
A. Queijeiro, Phys. Rev. {\bf D33}, 3449 (1986).
\bibitem{BERGMANN}L. Bergmann, Ph.D. thesis, Florida State University,
FSU-HEP-890215 (unpublished).
\bibitem{BHZ}V. Barger, T. Han and D. Zeppenfield, 
Phys. Rev. {\bf D41}, 2782 (1990).
\bibitem{STIRLING}J. Ohnemus and W. J. Stirling, 
Phys. Lett. {\bf B298}, 230 (1993).
\bibitem{HMRS}P. Harriman, A. Martin, R. Roberts, W. Stirling, 
Phys. Rev {\bf D42}, 798 (1990). 
\bibitem{NOTE}See, for example, Ref. \cite{BAUR} for a 
discussion of cuts to reduce the QCD contributions.
\end{references}
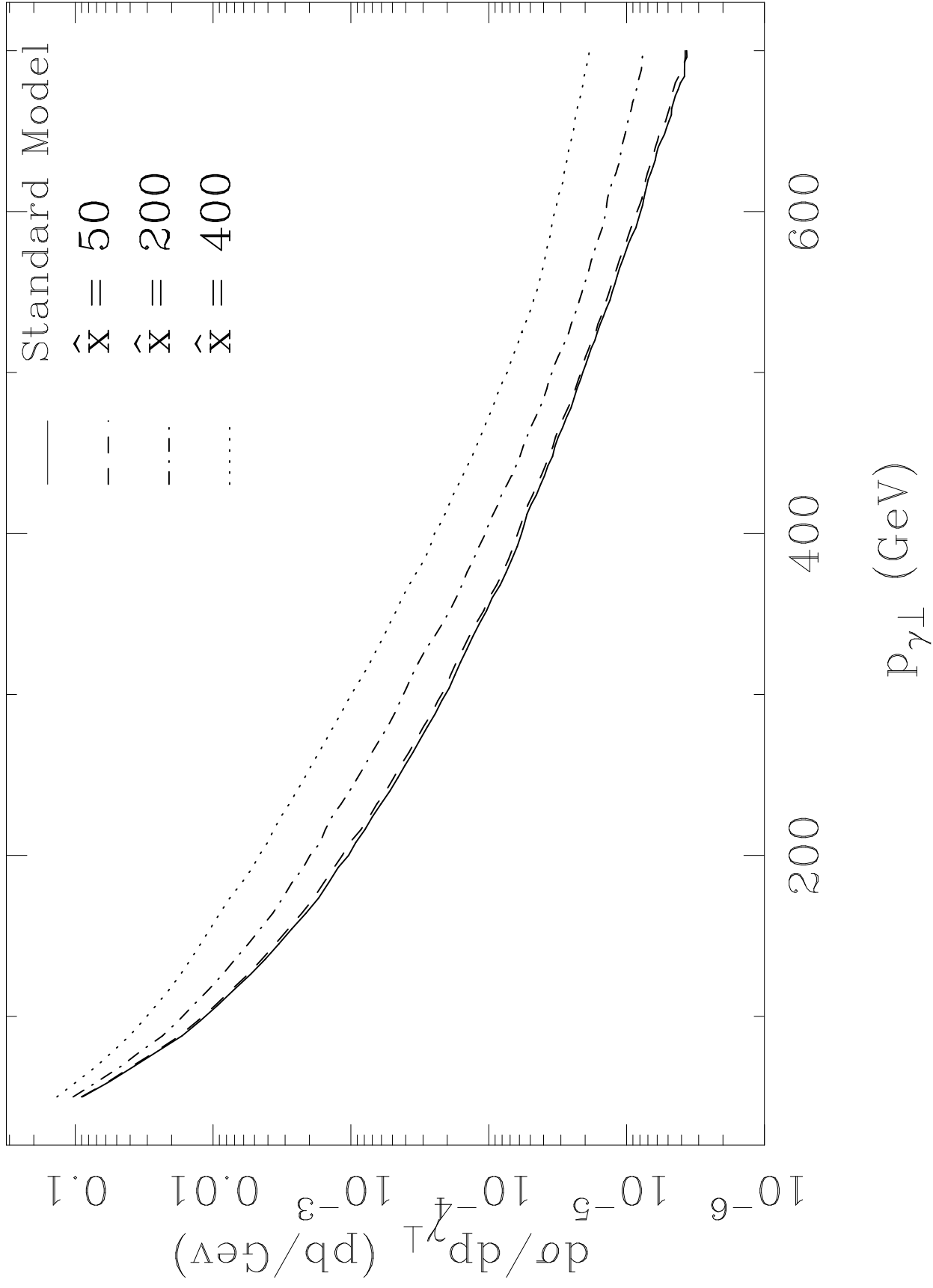
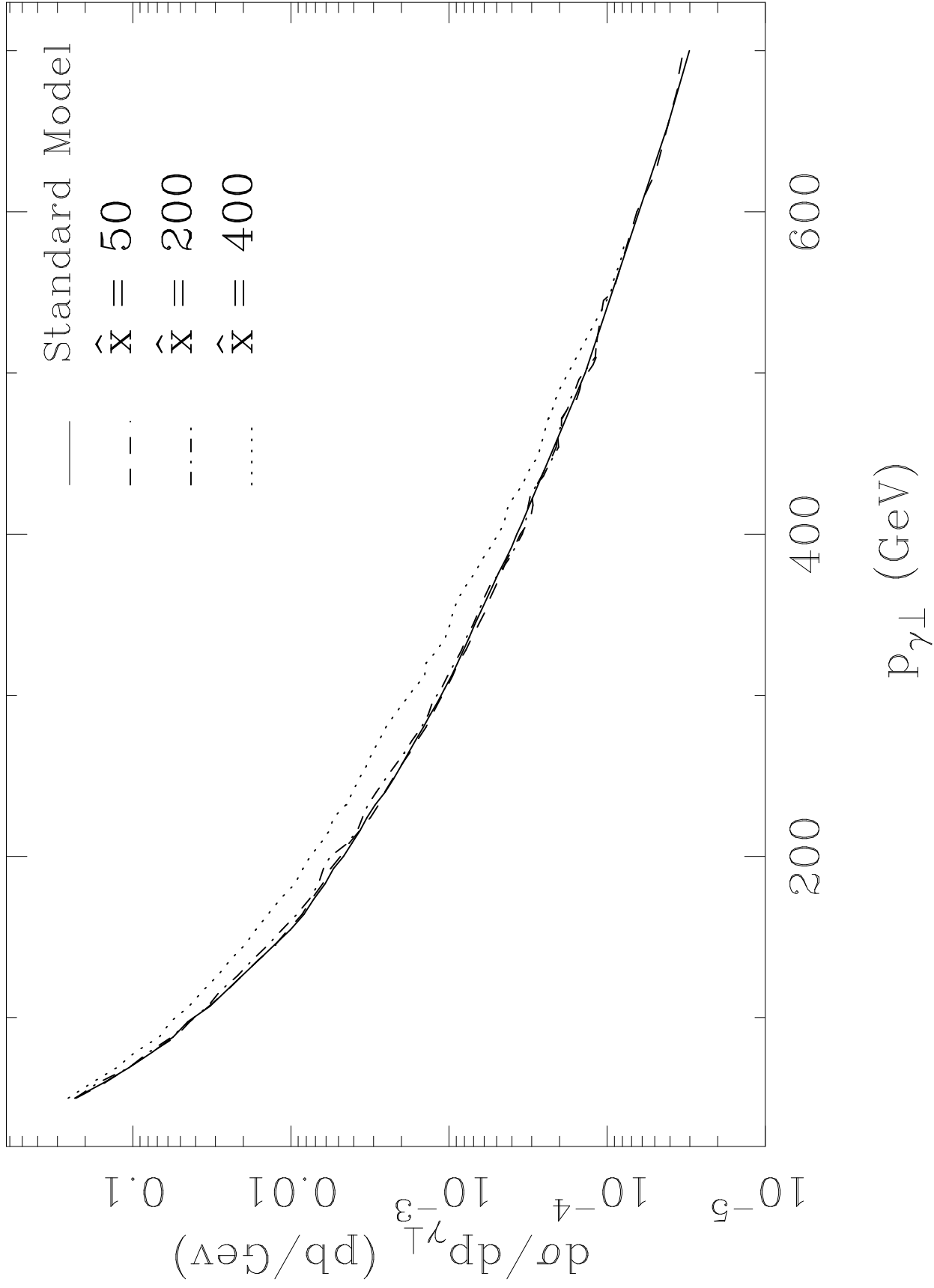
\end{document}